\documentclass{article}
\usepackage{dcase2023,amsmath,amssymb,graphicx,times,booktabs,tabularx,color,glossaries,siunitx,pgfplots,multirow,balance}
\usepackage[hyphens]{url}
\usepackage{hyperref}
\usepackage[hyphenbreaks]{breakurl}
\def\UrlBreaks{\do\/\do-}

\newacronym{DCASE}{DCASE}{Detection and Classification of Acoustic Scenes and Events}
\newacronym{STFT}{STFT}{short-time Fourier transform}
\newacronym{PSD}{PSD}{polyphonic sound detection}
\newacronym{PSDS}{PSDS}{polyphonic sound detection score}
\newacronym{PSDROC}{PSD-ROC}{PSD receiver operating characteristic}
\newacronym{piPSDS}{piPSDS}{post-processing independent PSDS}
\newacronym{piPSDROC}{pi-PSD-ROC}{pi-PSD-ROC}
\newacronym{miPSDS}{miPSDS}{median filter independent PSDS}
\newacronym{miPSDROC}{mi-PSD-ROC}{mi-PSD-ROC}
\newacronym{SED}{SED}{sound event detection}
\newacronym{TP}{TP}{true positive}
\newacronym{TPR}{TPR}{TP rate}
\newacronym{FP}{FP}{false positive}
\newacronym{FPR}{FPR}{FP rate}
\newacronym{eFPR}{eFPR}{effective FP rate}
\newacronym{TN}{TN}{true negative}
\newacronym{FN}{FN}{false negative}
\newacronym{CT}{CT}{cross triggers}
\newacronym{CTR}{CTR}{CT rate}
\title{Post-Processing Independent Evaluation of Sound Event Detection Systems}

\twoauthors
{Janek Ebbers, Reinhold Haeb-Umbach\thanks{Funded by the Deutsche Forschungsgemeinschaft (DFG, German Research Foundation) - 282835863.}}
{Paderborn University,\\
	Department of Communications Engineering,\\
	33098 Paderborn, Germany,\\\{ebbers,haeb\}@nt.upb.de}
{Romain Serizel}
{Universit\'{e} de Lorraine, CNRS,\\
	Inria, Loria,\\
	F-54000 Nancy, France,\\
	romain.serizel@loria.fr}

\begin{document}
	
	\ninept
    \setlength{\abovedisplayskip}{4pt}
    \setlength{\belowdisplayskip}{4pt}
    \setlength{\abovedisplayshortskip}{2pt}
    \setlength{\belowdisplayshortskip}{2pt}
    \setlength{\textfloatsep}{5pt}
	\maketitle
	
	\begin{sloppy}
		
		\begin{abstract}
			Due to the high variation in the application requirements of \gls{SED} systems, it is not sufficient to evaluate systems only in a single operating point.
			Therefore, the community recently adopted the \gls{PSDS}\glsunset{PSDROC} as an evaluation metric, which is the normalized area under the \gls{PSDROC}.
			It summarizes the system performance over a range of operating points.
			Hence, it provides a more complete picture of the overall system behavior and is less biased by hyper parameter tuning.
			So far \gls{PSDS} has only been computed over operating points resulting from varying the decision threshold that is used to translate the system output scores into a binary detection output.
			However, besides the decision threshold there is also the post-processing that can be changed to enter another operating mode.
			In this paper we propose the \gls{piPSDS} which computes \gls{PSDS} over operating points with varying post-processings and varying decision thresholds.
			It summarizes even more operating modes of an \gls{SED} system and allows for system comparison without the need of implementing a post-processing and without a bias due to different post-processings.
			While \gls{piPSDS} can in principle also combine different types of post-processing, we here, as a first step, present \gls{miPSDS} results for this year's DCASE Challenge Task4a systems.
			Source code is publicly available in our \text{sed\_scores\_eval} package\footnote{\url{https://github.com/fgnt/sed_scores_eval}}.
		\end{abstract}
		

        \vspace{-1mm}
		\section{Introduction}
        \vspace{-2mm}
		\label{sec:intro}
        \glsresetall
        Machine listening is recently attracting increased interest not only from academia but also from industry.
        It is the field of developing machines which can replicate the human ability of recognizing and processing a large number of different sounds.
        There are many sub-disciplines to machine listing, with \gls{SED}~\cite{mesaros2021sound} being one of them.
        Its aim is to recognize, classify and temporally localize sounds within an input audio.
        Due to the large number of possible applications, sounds and environments, one particular challenge is that there is often no or only little training data that perfectly matches the target application.
        Therefore, there is a particular interest in approaches for model training which can exploit imperfect data, such as weakly labeled learning~\cite{wang2019comparison,miyazaki2020weakly} and/or training with mismatched or unlabeled data~\cite{Lu2018,Ebbers2022}, as investigated by the \gls{DCASE} Challenge Task 4~\cite{dcase2023task4a} for several years now.

        Another more fundamental challenge for successful \gls{SED} system development is the meaningful evaluation and comparison of system performance, where the choice of the evaluation metric can have a large impact~\cite{ferroni2021improving}.
        Firstly, there is the complexity of the event matching between detected and ground truth events.
        Currently there exist three different approaches namely segment-based, collar-based and intersection-based~\cite{mesaros2016metrics,bilen2020framework}.
        The \gls{DCASE} Challenge Task 4 recently moved to intersection-based evaluation as it is more robust w.r.t. ambiguities in the ground truth labeling.
        Secondly, due to the high variation in application requirements, there is often not a single optimal system behavior as, e.g., expressed by the $F_1$-score.
        In some applications, missed hits may, e.g., be much more severe than false alarms.
        Therefore, system evaluation must ideally represent all different operating modes equally to capture the overall system behavior.
        The \gls{PSDS}~\cite{bilen2020framework,ebbers2022threshold} has been employed to capture performance over the range of decision thresholds, which are used to translate soft system output scores\footnote{Note the ambiguity of the term score here, where PSD score refers to a metric value while output scores refer to soft class activity predictions of a model/neural network.} into binary decisions.
        Therefore, system comparison using \gls{PSDS} is also less biased by threshold tuning w.r.t. to a certain operating point.

        However, there is also the post-processing that has a large impact on the system performance, which is mostly underinvestigated.
        In particular, system comparisons may be biased due to the employment of different post-processings.
        Similar to the decision threshold, the type and parameters of the post-processing can be understood as operating parameters of the system and may be adjusted to enter another operating mode.

        In this paper we propose \gls{piPSDS} which summarizes performance over both different post-processings and decision thresholds.
        Hence, it gives an even more complete picture of the system's performance over different operating modes and furthermore is less biased by hyper-parameter tuning.
        We perform investigations on this year's DCASE Challenge Task 4 submissions and show that 1) there is indeed a large impact on evaluation results due to post-processing 2) for different operating points there are different optimal post-processings and 3) the proposed \gls{piPSDS}\glsunset{piPSDROC} allows \gls{SED} system evaluation unbiased from threshold and post-processing tuning.

        The rest of the paper is structured as follows.
        First, we recapitulate the preliminaries of \gls{SED}, its evaluation and the \gls{PSDS} in Sec.~\ref{sec:sed}, Sec.~\ref{sec:sedeval} and Sec.~\ref{sec:psds}, respectively.
        Our proposed \gls{piPSDS} is presented in Sec.~\ref{sec:pipsds}.
        Finally, we show results in Sec.~\ref{sec:results} and draw conclusions in Sec.~\ref{sec:conclusions}.

		\section{Preliminaries}
        \vspace{-1mm}
		\subsection{Sound Event Detection}
        \vspace{-1mm}
		\label{sec:sed}
		To not only recognize but also temporally localize sound events, \gls{SED} systems perform multi-label classification within smaller time-windows of an audio clip, e.g., at \gls{STFT} frame-level.
		For each window $n$ a system provides soft classification scores $y_{n,c}$ for each event class $c$ out of a set of $C$ predefined sound event classes of interest.
		These scores represent the predicted activity of the event within a particular time-window.
		To obtain a hard decision, soft classification scores can be binarized using a certain decision threshold $\gamma_c$, where the class $c$ is assumed active in the \text{$n$-th} window if $y_{n,c} \geq \gamma_c$, else it is assumed inactive.
		Connected active windows are then merged into a detected event $(\hat{t}_{\text{on},i}, \hat{t}_{\text{off},i}, \hat{c}_i)$ defined by onset time $\hat{t}_{\text{on},i}$, offset time $\hat{t}_{\text{off},i}$ and class label $\hat{c}_i$, respectively, where $i$ represents the event index.
		Usually it is beneficial to run some kind of post-processing before or after binariazation to obtain meaningfully connected event predictions and be more robust w.r.t. outliers.
		The type and hyper-parameters of the post-processing, as well as the decision threshold and any other hyper-parameters that may be easily changed during application are summarized as a system's operating parameters $\tau$ in the following.
		
		\subsection{Evaluation of Detected Events}
        \vspace{-1mm}
		\label{sec:sedeval}
		The evaluation of the detected events of event class $c$ for specific operating parameters~$\tau$ is, in accordance with other classification tasks, based on counting the intermediate statistics $N_{\text{TP},c,\tau}$, $N_{\text{FN},c,\tau}$ and $N_{\text{FP},c,\tau}$, which refer to the numbers of
		\vspace{-1mm}
		\begin{itemize}
			\itemsep0mm 
			\item ground truth events that have been correctly detected by the system a.k.a. \gls{TP} detections,
			\item ground truth events that have not been detected by the system a.k.a. \gls{FN} detections,
			\item detected events that do not match any ground truth event a.k.a. \gls{FP} detections,
		\end{itemize}
		\vspace{-1.8mm}
		accumulated over the whole evaluation set, respectively.
		Bilen et al.~\cite{bilen2020framework} have further taken \glspl{CT} into account, a.k.a. substitutions, with $N_{\text{CT},c,k,\tau}$ being the number of \glspl{FP} of class $c$ matching ground truth events from another event class $k$, which may impair user experience more than standalone \glspl{FP}.
		
		For counting above intermediate statistics there exist different approaches w.r.t. how the temporal matching between detected events and ground truth events is handled.
		As the definitions of \gls{PSDS} and \gls{piPSDS}\glsunset{PSDROC}\glsunset{piPSDROC}, however, do not depend on the temporal matching that is used, we here only briefly recapitulate intersection-based evaluation which is the approach recently used for \gls{PSDS} computation as it is more robust w.r.t. ambiguities in the labeling of the evaluation data.
		Note, however, that one could instead also compute segment-based and collar-based~\cite{mesaros2016metrics} \gls{PSDS}/ \gls{piPSDS}.
		
		Intersection-based evaluation requires detected events to intersect with ground truth events by at least $\rho_\text{DTC}$ to be not counted as a \gls{FP} detection.
		Moreover, it requires a ground truth event to intersect with non-\gls{FP} events by at least $\rho_\text{GTC}$ to be counted as a \gls{TP} detection.
		Further, if an \gls{FP} event intersects with a ground truth event of another class by at least $\rho_\text{CTTC}$ it is counted as a \gls{CT}. 
		
		Of particular interest are in the following the \gls{TPR} defined as $r_{c,\tau}=\frac{N_{\text{TP},c,\tau}}{N_{\text{TP},c,\tau}+N_{\text{FN},c,\tau}}$, and the \gls{eFPR}\glsunset{FPR}
		\begin{align}
			e_{c,\tau} = \frac{N_{\text{FP},c,\tau}}{T_\text{ds}} + \alpha_\text{CT}\frac{1}{C-1}\sum_{\substack{k\\k\neq c}}\frac{N_{\text{CT},c,k,\tau}}{T_k}\,\,\, .
		\end{align}
		which consists of the \gls{FPR} $\frac{N_{\text{FP},c,\tau}}{T_\text{ds}}$ plus an additional penalty on \glspl{CTR} $\frac{N_{\text{CT},c,k,\tau}}{T_k}$ averaged over all other classes $k\neq c$ and weighted by $\alpha_\text{CT}$.
		Note that, with intersection-based evaluation, there is not a countable number of negative events, which is why the \gls{FPR} is computed w.r.t. the total duration of the evaluation dataset $T_\text{ds}$, whereas \glspl{CTR} are computed w.r.t. the total duration of activity $T_k$ of the $k$-th class within the evaluation dataset.
		
		\subsection{Polyphonic Sound Detection Score}
        \vspace{-1mm}
		\label{sec:psds}
		To compute \gls{PSDS}~\cite{bilen2020framework}, one starts with the computation of single-class \gls{PSDROC} curves $r_c(e)$ for each event class $c$.
		$r_c(e)$ is obtained as a continuous "staircase-type" interpolation of true positive rates $r_{c,\tau}$ plotted over corresponding \glspl{eFPR} $e_{c,\tau}$ for different operating parameters $\tau\in\widehat{\mathcal{T}}_c$.
		
		While $\tau$ may be any (set of) hyper-parameter(s) that may change system behavior, it has so far, in accordance with the standard definition of ROC curves~\cite{Davis06therelationship}, only been considered to be the decision threshold used to translate soft prediction scores into binary detections.
		Here, an  algorithm for the efficient joint evaluation of all possible decision thresholds has been proposed in~\cite{ebbers2022threshold}.
		Note that, in contrast to standard ROC curves, it is here not always guaranted that $r_{c,\tau}$ is monotonically increasing with $e_{c,\tau}$, when, e.g., sophisticated intersection-based evaluation is employed.
		As in operation, however, one would always prefer the operating point with a higher true positive rate at lower or equal false positive rate if available, $\widehat{\mathcal{T}}_c$ represents only best case operating parameters:
		\begin{align}
			\widehat{\mathcal{T}}_c = \big\{ \tau \big|\,\nexists\,\lambda\,\text{with}\,e_{c,\lambda}\leq e_{c,\tau}\, \text{and}\, r_{c,\lambda}>r_{c,\tau}  \big\}.\label{eq:best_case_ops}
		\end{align}
		
		Having the single-class PSD-ROC curves $r_c(e)$, the overall PSD-ROC curve is defined as the effective true positive rate 
		\begin{align}
			r(e) = \mu_\text{TP}(e)  - \alpha_\text{ST} \sigma_\text{TP}(e)\label{eq:piPSDROC}
		\end{align}
		which is average per-class true positive rate minus a penalty on standard deviation over classes weighted by a metric parameter $\alpha_\text{ST}$ with
		\begin{align*}
			\mu_\text{TP}(e)=\frac{1}{C}\sum_{c=1}^{C} r_c(e);\quad
			\sigma_\text{TP}(e)=\sqrt{\frac{1}{C}\sum_{c=1}^{C} (r_c(e) - \mu_\text{TP}(e))^2}.
		\end{align*}
		
		Finally, the \gls{PSDS} is the normalized area under the \gls{PSDROC}:
		\begin{align}
			\text{PSDS} = \frac{1}{e_\text{max}}\int_0^{e_\text{max}}r(e)de\label{eq:psds}
		\end{align}
		with the maximal false positive rate $e_\text{max}$ being a metric parameter, which controls up to which false positive rate the operating points may still be relevant.

		\section{Post-Processing Independent Polyphonic Sound Detection Score}
        \vspace{-1mm}
		\label{sec:pipsds}
		
		Besides the decision threshold there is also the post-processing that we could change to enter another operating mode.
		As an example, Fig.~\ref{fig:psd_rocs_speech} shows the single-class PSD-ROC curves for "Speech" from this year's "Baseline\_BEATS" system~\cite{baseline2023beats} when using post-processing median filtering with lengths of $\SI{0.1}{s}$ and $\SI{1.0}{s}$, respectively.
		It appears that when the system is operated in low eFPR mode, than it is better to use the larger median filter window size.
		When the system should be operated in high TPR mode, it is better to use a smaller window size.
		Thus, it is reasonable and also fairly easy to choose the post-processing depending on the requirements of a given application.
		To account for this in the system evaluation, which is supposed to capture overall system behavior, we propose to incorporate the variation of post-processing into the computation of the \gls{PSDS} to get a \glsreset{piPSDS}\gls{piPSDS}.

		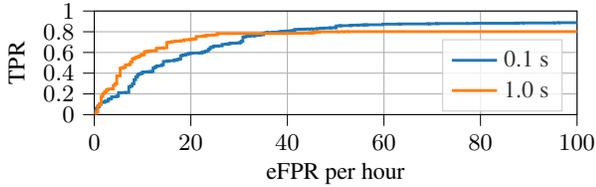
\begin{figure}[t]
			\centering
			\newlength\figureheight
			\newlength\figurewidth
			\setlength\figureheight{2.96cm}
			\setlength\figurewidth{8cm}
\begin{tikzpicture}

\definecolor{darkgray176}{RGB}{176,176,176}
\definecolor{darkorange25512714}{RGB}{255,127,14}
\definecolor{lightgray204}{RGB}{204,204,204}
\definecolor{steelblue31119180}{RGB}{31,119,180}

\begin{axis}[
height=\figureheight,
width=\figurewidth,
legend cell align={left},
legend style={
  fill opacity=0.8,
  draw opacity=1,
  text opacity=1,
  at={(0.97,0.03)},
  anchor=south east,
  draw=lightgray204
},
tick align=outside,
tick pos=left,
x grid style={darkgray176},
xlabel={eFPR per hour},
xmajorgrids,
xmin=0, xmax=100,
xtick style={color=black},
y grid style={darkgray176},
ylabel={TPR},
y label style={yshift=-2mm},
ymajorgrids,
ymin=0, ymax=1,
ytick style={color=black}
]
\addplot [line width=1.08pt, steelblue31119180, const plot mark left]
table {%
0 0.0390469887491727
0.357988106839562 0.0615486432825943
0.715976213679123 0.0761085373924553
1.07396432051869 0.101919258769027
1.43195242735825 0.117802779616148
2.14792864103737 0.129053606882859
2.50591674787693 0.129715420251489
2.86390485471649 0.14824619457313
3.22189296155606 0.152217074784911
3.57988106839562 0.171409662475182
3.93786917523518 0.172071475843812
4.6538453889143 0.179351422898743
5.01183349575386 0.211780277961615
5.36982160259343 0.212442091330245
6.44378592311211 0.214427531436135
7.15976213679123 0.268034414295169
7.5177502436308 0.277299801455989
7.87573835047036 0.297154202514891
8.23372645730992 0.337524818001324
8.59171456414948 0.363335539377895
8.94970267098904 0.386499007279947
9.3076907778286 0.395102581072138
9.66567888466817 0.398411647915288
10.0236669915077 0.411647915287889
10.3816550983473 0.412971542025149
11.0976313120264 0.413633355393779
11.455619418866 0.415618795499669
11.8136075257055 0.416280608868299
12.1715956325451 0.449371277299801
12.5295837393847 0.458636664460622
12.8875718462242 0.469887491727333
13.2455599530638 0.471872931833223
13.9615361667429 0.481138318994044
14.3195242735825 0.513567174056916
15.0355004872616 0.515552614162806
15.3934885941012 0.520185307743216
16.4674529146198 0.524156187954997
16.8254410214594 0.528788881535407
17.183429128299 0.532759761747187
17.5414172351385 0.536068828590338
17.8994053419781 0.56783587028458
18.2573934488176 0.57445400397088
18.6153815556572 0.57776307081403
18.9733696624968 0.581733951025811
19.3313577693363 0.587690271343481
19.6893458761759 0.590337524818001
20.0473339830155 0.594308405029782
20.405322089855 0.594970218398412
21.4792864103737 0.595632031767042
21.8372745172133 0.596293845135672
22.1952626240528 0.598941098610192
22.5532507308924 0.605559232296492
22.911238837732 0.610191925876903
23.2692269445715 0.615486432825943
23.6272150514111 0.631369953673064
23.9852031582506 0.647915287888815
24.3431912650902 0.652547981469226
24.7011793719298 0.661813368630046
25.0591674787693 0.662475181998676
25.4171555856089 0.663136995367306
25.7751436924484 0.663798808735936
26.4911199061276 0.678358702845797
27.2070961198067 0.679682329583058
27.5650842266463 0.682329583057578
27.9230723334858 0.686962276637988
28.2810604403254 0.688285903375248
28.6390485471649 0.689609530112508
29.3550247608441 0.691594970218398
30.0710009745232 0.695565850430179
30.4289890813627 0.70880211780278
30.7869771882023 0.749172733289212
31.1449652950419 0.752481800132363
31.5029534018814 0.755790866975513
31.860941508721 0.757114493712773
32.2189296155606 0.760423560555923
32.9349058292397 0.761085373924553
33.2928939360792 0.764394440767703
33.6508820429188 0.766379880873594
34.0088701497584 0.767703507610854
34.3668582565979 0.771012574454004
35.082834470277 0.786896095301125
35.4408225771166 0.790205162144275
35.7988106839562 0.792190602250165
36.1567987907957 0.792852415618795
36.8727750044749 0.793514228987426
37.2307631113144 0.796161482461946
37.9467393249935 0.797485109199206
38.3047274318331 0.800132362673726
38.6627155386727 0.801455989410986
39.0207036455122 0.802117802779616
39.3786917523518 0.804765056254136
39.7366798591914 0.806750496360026
40.0946679660309 0.807412309728657
40.4526560728705 0.808074123097287
40.81064417971 0.809397749834547
41.1686322865496 0.810059563203177
41.5266203933892 0.823295830575778
42.6005847139078 0.824619457313038
43.316560927587 0.825281270681668
43.6745490344265 0.825943084050298
44.3905252481056 0.826604897418928
45.1065014617848 0.827266710787558
46.1804657823035 0.831237590999338
46.8964419959826 0.833223031105228
47.2544301028221 0.834546657842488
47.6124182096617 0.835870284579749
48.6863825301804 0.837855724685639
49.04437063702 0.838517538054269
49.7603468506991 0.839179351422899
50.1183349575386 0.85638649900728
50.8343111712178 0.85837193911317
51.1922992780573 0.8590337524818
52.9822398122551 0.85969556585043
53.3402279190947 0.86366644606221
54.4141922396134 0.86432825943084
54.7721803464529 0.864990072799471
55.4881565601321 0.865651886168101
55.8461446669716 0.866313699536731
56.2041327738112 0.866975512905361
56.5621208806507 0.867637326273991
59.7840138422068 0.873593646591661
63.0059068037629 0.874917273328921
63.3638949106024 0.875579086697551
69.0917046200354 0.876902713434811
69.449692726875 0.877564526803441
70.8816451542332 0.878226340172071
71.2396332610728 0.879549966909332
77.3254310773453 0.880211780277962
78.7573835047036 0.881535407015222
79.1153716115431 0.882197220383852
86.9911099620135 0.882859033752482
87.3490980688531 0.883520847121112
87.7070861756926 0.884182660489742
88.7810504962113 0.884844473858372
95.2248364193234 0.886168100595632
96.2988007398421 0.887491727332892
100 0.887491727332892
};
\addlegendentry{0.1 s}
\addplot [line width=1.08pt,  darkorange25512714, const plot mark left]
table {%
0 0.0132362673726009
0.357988106839562 0.0271343481138319
0.715976213679123 0.0866975512905361
1.07396432051869 0.0992720052945069
1.43195242735825 0.176704169424222
1.78994053419781 0.203838517538054
2.14792864103737 0.219722038385175
2.50591674787693 0.239576439444077
2.86390485471649 0.247518199867637
3.57988106839562 0.248180013236267
3.93786917523518 0.2819324950364
4.29585728207474 0.297816015883521
4.6538453889143 0.373924553275976
5.36982160259343 0.448047650562541
5.72780970943299 0.452018530774322
6.08579781627255 0.457313037723362
6.44378592311211 0.480476505625414
7.15976213679123 0.485109199205824
7.5177502436308 0.508934480476506
7.87573835047036 0.538716082064858
8.23372645730992 0.542686962276638
8.59171456414948 0.553937789543349
8.94970267098904 0.557908669755129
9.3076907778286 0.56849768365321
9.66567888466817 0.57842488418266
10.0236669915077 0.57908669755129
10.3816550983473 0.607544672402383
10.7396432051869 0.608206485771013
11.0976313120264 0.610191925876903
11.455619418866 0.615486432825943
11.8136075257055 0.616148246194573
12.5295837393847 0.617471872931833
12.8875718462242 0.641297154202515
13.9615361667429 0.643282594308405
14.3195242735825 0.645268034414295
15.0355004872616 0.694904037061549
15.3934885941012 0.699536730641959
15.7514767009407 0.701522170747849
16.1094648077803 0.705493050959629
16.8254410214594 0.70814030443415
17.183429128299 0.70946393117141
17.5414172351385 0.71012574454004
17.8994053419781 0.71409662475182
18.6153815556572 0.722700198544011
19.6893458761759 0.724023825281271
20.0473339830155 0.726009265387161
20.405322089855 0.729318332230311
20.7633101966946 0.732627399073461
21.1212983035341 0.754467240238253
22.5532507308924 0.760423560555923
22.911238837732 0.761085373924553
23.2692269445715 0.764394440767703
23.6272150514111 0.767041694242224
23.9852031582506 0.767703507610854
24.3431912650902 0.768365320979484
24.7011793719298 0.769027134348114
25.4171555856089 0.782263401720715
25.7751436924484 0.782925215089345
39.7366798591914 0.784248841826605
41.5266203933892 0.785572468563865
42.9585728207474 0.786896095301125
44.3905252481056 0.788881535407015
44.7485133549452 0.789543348775645
45.1065014617848 0.790205162144275
45.4644895686243 0.792852415618795
45.8224776754639 0.794837855724686
46.1804657823035 0.798146922567836
46.538453889143 0.799470549305096
50.4763230643782 0.800132362673726
51.5502873848969 0.800794176042356
52.6242517054156 0.801455989410986
100 0.801455989410986
};
\addlegendentry{1.0 s}
\end{axis}

\end{tikzpicture}
			\vspace{-3mm}
			\caption{Baseline Speech ROCs with different median filters}
			\label{fig:psd_rocs_speech}
		\end{figure}
		
		To do so, we consider the operating parameters $\tau=(l, \gamma)$ to be a tuple of the post-processing~$~l$ and the decision threshold~$\gamma$.
		Here, $l$ defines which postprocessing is used out of a predefined set of $L$ possible post-processings.
		Post-processing is assumed to be applied before the binarization with the decision threshold $\gamma$, as this still allows to use the algorithm from~\cite{ebbers2022threshold} to efficiently compute the true positive and \glspl{eFPR} for the whole continuous range of $\gamma$.
		Here, it is worth noting that, e.g., median filtering and thresholding are permutation invariant, i.e., applying the median filter before binarization yields the same result as applying it afterwards.

		The definition of the \gls{PSDROC} according to Sec.~\ref{sec:psds} with $\tau \in \mathbb{L} \times \mathbb{R}$, where \text{$\mathbb{L}=\{l\in\mathbb{N}\,|\,l < L \}$}, gives us the \gls{piPSDROC}.
		Due to the restriction to best case operating points in Eq.~\ref{eq:best_case_ops} the single-class \glspl{piPSDROC} can be computed as
        \vspace{-0.5mm}
		\begin{align}
			r_c(e) = \max_l{r_{c,l}(e)}
		\end{align}
        \vspace{-1mm}
		where $r_{c,l}(e)$ represents the single-class \gls{PSDROC} for a single post-processing $l$ resulting only from variation of the decision threshold.
		Hence, the single-class \gls{piPSDROC} chooses, for a given eFPR $e$, the post-processing with the highest TPR.
		\gls{piPSDS} is then, analogously to Eq.~\ref{eq:psds}, the normalized area under the \gls{piPSDROC}.

		Overall, \gls{piPSDS} has two major advantages over only threshold-independent \gls{PSDS}. Firstly, it better captures real-world \gls{SED} applications, where it is natural to choose the post-processing that best suits the current application requirements. Secondly, for research it allows for system comparison without a bias being introduced by different post-processings.

        \vspace{-1mm}
		\section{Results}
        \vspace{-1mm}
		\label{sec:results}
		Investigations are done with the baseline and submissions of this year's DCASE challenge Task4a.
		Participants have been asked to, in addition to their post-processed submission, also share the raw prediction scores as provided by their model/neural network without any further post-processing.
		This allows us to investigate 1) the impact of the post-processing, 2) post-processing independent evaluation.
		All following evaluations are performed on the DESED~\cite{turpault2019sound} public eval set, which is a part of the challenge evaluation data.

		There are two intersection-based PSDS evaluated in the challenge, which refer to different scenarios.
		PSDS1 ($\rho_\text{DTC}=0.7$, $\rho_\text{GTC}=0.7$, $\alpha_\text{CT}=0$, $\alpha_\text{ST}=1$, $e_\text{max}=100/\text{hour}$) particularly evaluates the model's capability of temporally localizing sound events, whereas PSDS2 ($\rho_\text{DTC}=0.1$, $\rho_\text{GTC}=0.1$, $\rho_\text{CTTC}=0.3$, $\alpha_\text{CT}=0.5$, $\alpha_\text{ST}=1$, $e_\text{max}=100/\text{hour}$) is more focused on evaluating the reliable recognition of event classes within an audio clip.
		Due to space constraints and with post-processing being particularly relevant for the temporal localization of sound events, we only consider PSDS1 evaluation in the following.

		With median filtering being the most popular type of post-processing for \gls{SED} systems, we here consider \gls{miPSDS} as an instance of \gls{piPSDS}, where the set of possible post-processings consists of median filters with different filter lengths.
		As the set of median filter lengths we use 21 filter lengths linearly spaced from \SI{0.0}{s} (no filtering) to \SI{1.0}{s}, 10 from \SI{1.1}{s} to \SI{2.0}{s}, 5 from \SI{2.2}{s} to \SI{3.0}{s} and 4 from \SI{3.5}{s} to \SI{5.0}{s} overall totaling 40 different filter lengths.
		The implementation of the median filter equals a time continuous filtering of a piece-wise constant signal that is defined by the timestamped prediction scores submitted by the participants.
		This way it is ensured the systems employ the very same post-processing regardless of the system's output resolution which may vary across systems.
		Implementations of the median filter, \gls{miPSDS} and \gls{piPSDS}, with the latter taking any list of differently post-processed scores, are publicly available in the \text{sed\_scores\_eval} package\footnotemark[1], that is, in accordance with the challenge, used for evaluation.

		\begin{figure}[t]
			\centering
			\setlength\figureheight{3.3cm}
			\setlength\figurewidth{8cm}
			\input{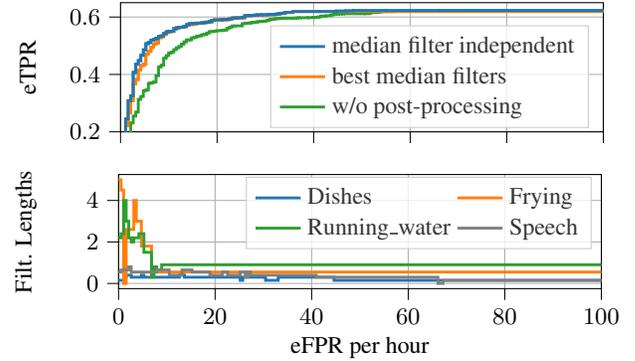}
			\vspace{-1mm}
			\setlength\figureheight{3.1cm}
			\setlength\figurewidth{8cm}
\begin{tikzpicture}

\definecolor{darkgray176}{RGB}{176,176,176}
\definecolor{darkorange25512714}{RGB}{255,127,14}
\definecolor{forestgreen4416044}{RGB}{44,160,44}
\definecolor{gray127}{RGB}{127,127,127}
\definecolor{lightgray204}{RGB}{204,204,204}
\definecolor{steelblue31119180}{RGB}{31,119,180}

\begin{axis}[
height=\figureheight,
width=\figurewidth,
legend cell align={left},
legend style={fill opacity=0.8, draw opacity=1, text opacity=1, draw=lightgray204},
legend columns=2,
tick align=outside,
tick pos=left,
x grid style={darkgray176},
xlabel={eFPR per hour},
xmajorgrids,
xmin=0, xmax=100,
xtick style={color=black},
y grid style={darkgray176},
ylabel={Filt. Lengths},
ymajorgrids,
ymin=-0.25, ymax=5.25,
ytick style={color=black}
]
\addplot [line width=1.08pt, steelblue31119180, const plot mark left]
table {%
0 0.15
0.52498796902571 0.15
1.04997593805142 0.3
1.57496390707713 0.4
2.09995187610284 0.4
2.62493984512855 0.3
3.14992781415426 0.3
4.19990375220568 0.3
4.72489172123139 0.4
5.2498796902571 0.3
7.34983156635994 0.3
7.87481953538565 0.3
8.39980750441136 0.3
8.92479547343707 0.3
9.44978344246278 0.3
9.97477141148849 0.3
11.0247473495399 0.3
11.5497353185656 0.3
12.599711256617 0.4
13.6496871946685 0.3
14.1746751636942 0.3
15.7496390707713 0.3
16.7996150088227 0.3
17.8495909468741 0.3
18.3745789158998 0.3
18.8995668849256 0.3
19.4245548539513 0.3
22.5744826681055 0.3
23.0994706371312 0.3
25.1994225132341 0.15
25.7244104822598 0.3
28.3493503273883 0.3
30.4493022034912 0.15
32.0242661105683 0.15
32.549254079594 0.15
33.0742420486197 0.3
33.5992300176454 0.3
34.1242179866711 0.3
35.6991818937483 0.3
44.6239773671853 0.15
46.7239292432882 0.15
100 0.15
};
\addlegendentry{{\small Dishes}}
\addplot [line width=1.08pt, darkorange25512714, const plot mark left]
table {%
0 5
0.52498796902571 4.5
1.04997593805142 0
1.57496390707713 2.6
3.14992781415426 4
3.67491578317997 3
4.72489172123139 1.8
6.82484359733423 0.55
100 0.55
};
\addlegendentry{{\small Frying}}
\addplot [line width=1.08pt, forestgreen4416044, const plot mark left]
table {%
0 2.2
0.52498796902571 2.4
1.04997593805142 4
1.57496390707713 3
2.09995187610284 2.2
2.62493984512855 2
3.14992781415426 2.2
4.19990375220568 2.4
5.2498796902571 1.5
6.29985562830852 1.5
6.82484359733423 0.3
7.34983156635994 0.8
7.87481953538565 0.65
8.92479547343707 0.9
100 0.9
};
\addlegendentry{{\small Running\_water}}
\addplot [line width=1.08pt, gray127, const plot mark left]
table {%
0 0.55
0.52498796902571 0.65
1.04997593805142 0.65
1.57496390707713 0.65
2.09995187610284 0.8
2.62493984512855 0.55
3.14992781415426 0.55
3.67491578317997 0.55
4.19990375220568 0.55
4.72489172123139 0.55
5.2498796902571 0.55
5.77486765928281 0.55
6.29985562830852 0.55
6.82484359733423 0.55
7.34983156635994 0.55
7.87481953538565 0.65
8.39980750441136 0.65
8.92479547343707 0.65
9.44978344246278 0.65
10.4997593805142 0.4
11.0247473495399 0.4
11.5497353185656 0.4
12.0747232875913 0.4
12.599711256617 0.4
13.1246992256427 0.65
13.6496871946685 0.65
14.1746751636942 0.65
14.6996631327199 0.65
15.2246511017456 0.55
15.7496390707713 0.55
16.7996150088227 0.55
17.3246029778484 0.55
17.8495909468741 0.55
18.3745789158998 0.55
18.8995668849256 0.4
19.4245548539513 0.4
20.9995187610284 0.4
21.5245067300541 0.55
22.0494946990798 0.55
22.5744826681055 0.4
23.0994706371312 0.4
23.6244586061569 0.4
25.1994225132341 0.55
25.7244104822598 0.55
26.2493984512855 0.3
26.7743864203112 0.55
27.2993743893369 0.4
29.3993262654398 0.4
30.9742901725169 0.4
31.4992781415426 0.4
32.0242661105683 0.4
40.9490615840054 0.3
41.4740495530311 0.3
55.1237367476995 0.3
66.1484840972394 0
67.1984600352909 0.15
71.9233517565222 0.15
100 0.15
};
\addlegendentry{{\small Speech}}
\end{axis}

\end{tikzpicture}
			\vspace{-6mm}
			\caption{Upper Plot: PSD-ROCS for different post-processing setups. Lower  Plot: Optimal median filter lengths over operating points as tracked by median filter independent PSD-ROC.}
			\label{fig:baseline_rocs}
		\end{figure}

		\begin{figure*}[t]
			\centering
			\setlength\figureheight{3.7cm}
			\setlength\figurewidth{18cm}
\begin{tikzpicture}

\definecolor{darkgray176}{RGB}{176,176,176}
\definecolor{forestgreen4416044}{RGB}{44,160,44}
\definecolor{gray127}{RGB}{127,127,127}
\definecolor{lightgray204}{RGB}{204,204,204}

\begin{axis}[
height=\figureheight,
width=\figurewidth,
legend cell align={left},
legend style={fill opacity=0.8, draw opacity=1, text opacity=1, draw=lightgray204,at={(.999,.99)}},
tick align=outside,
tick pos=left,
x grid style={darkgray176},
xmajorgrids,
xmin=-0.966666666666667, xmax=12.9666666666667,
xtick style={color=black},
xtick={0,1,2,3,4,5,6,7,8,9,10,11},
xticklabels=\empty,
y grid style={darkgray176},
ylabel={PSDS1},
ymajorgrids,
ymin=0.33, ymax=0.7,
ytick style={color=black}
]
\draw[draw=none,fill=gray127] (axis cs:-0.333333333333333,0) rectangle (axis cs:0,0.645200126516441);
\addlegendimage{ybar,ybar legend,draw=none,fill=gray127}
\addlegendentry{\footnotesize w/ post-processing}

\draw[draw=none,fill=gray127] (axis cs:0.666666666666667,0) rectangle (axis cs:1,0.621141020788742);
\draw[draw=none,fill=gray127] (axis cs:1.66666666666667,0) rectangle (axis cs:2,0.60471918350143);
\draw[draw=none,fill=gray127] (axis cs:2.66666666666667,0) rectangle (axis cs:3,0.598379881090981);
\draw[draw=none,fill=gray127] (axis cs:3.66666666666667,0) rectangle (axis cs:4,0.593545783842825);
\draw[draw=none,fill=gray127] (axis cs:4.66666666666667,0) rectangle (axis cs:5,0.573239383095762);
\draw[draw=none,fill=gray127] (axis cs:5.66666666666667,0) rectangle (axis cs:6,0.571634859831974);
\draw[draw=none,fill=gray127] (axis cs:6.66666666666667,0) rectangle (axis cs:7,0.569203728123046);
\draw[draw=none,fill=gray127] (axis cs:7.66666666666667,0) rectangle (axis cs:8,0.560266812872252);
\draw[draw=none,fill=gray127] (axis cs:8.66666666666667,0) rectangle (axis cs:9,0.550987663690949);
\draw[draw=none,fill=gray127] (axis cs:9.66666666666667,0) rectangle (axis cs:10,0.47475822315399);
\draw[draw=none,fill=gray127] (axis cs:10.6666666666667,0) rectangle (axis cs:11,0.449871805537578);
\draw[draw=none,fill=gray127] (axis cs:11.6666666666667,0) rectangle (axis cs:12,0.426695730428801);
\draw[draw=none,fill=forestgreen4416044] (axis cs:-2.77555756156289e-17,0) rectangle (axis cs:0.333333333333333,0.610041021703561);
\addlegendimage{ybar,ybar legend,draw=none,fill=forestgreen4416044}
\addlegendentry{\footnotesize w/o post-processing}

\draw[draw=none,fill=forestgreen4416044] (axis cs:1,0) rectangle (axis cs:1.33333333333333,0.609431400778044);
\draw[draw=none,fill=forestgreen4416044] (axis cs:2,0) rectangle (axis cs:2.33333333333333,0.590559497123866);
\draw[draw=none,fill=forestgreen4416044] (axis cs:3,0) rectangle (axis cs:3.33333333333333,0.58307172182101);
\draw[draw=none,fill=forestgreen4416044] (axis cs:4,0) rectangle (axis cs:4.33333333333333,0.590804699639849);
\draw[draw=none,fill=forestgreen4416044] (axis cs:5,0) rectangle (axis cs:5.33333333333333,0.586390093018254);
\draw[draw=none,fill=forestgreen4416044] (axis cs:6,0) rectangle (axis cs:6.33333333333333,0.549406614636121);
\draw[draw=none,fill=forestgreen4416044] (axis cs:7,0) rectangle (axis cs:7.33333333333333,0.567529003039599);
\draw[draw=none,fill=forestgreen4416044] (axis cs:8,0) rectangle (axis cs:8.33333333333333,0.569420509153888);
\draw[draw=none,fill=forestgreen4416044] (axis cs:9,0) rectangle (axis cs:9.33333333333333,0.545819108570566);
\draw[draw=none,fill=forestgreen4416044] (axis cs:10,0) rectangle (axis cs:10.3333333333333,0.471207593947271);
\draw[draw=none,fill=forestgreen4416044] (axis cs:11,0) rectangle (axis cs:11.3333333333333,0.461402970846246);
\draw[draw=none,fill=forestgreen4416044] (axis cs:12,0) rectangle (axis cs:12.3333333333333,0.387522511862413);
\path [draw=black, semithick]
(axis cs:-0.166666666666667,0.621509672914657)
--(axis cs:-0.166666666666667,0.669767556742262);

\path [draw=black, semithick]
(axis cs:0.833333333333333,0.599949249621053)
--(axis cs:0.833333333333333,0.639387935663913);

\path [draw=black, semithick]
(axis cs:1.83333333333333,0.590733918699181)
--(axis cs:1.83333333333333,0.626576986183954);

\path [draw=black, semithick]
(axis cs:2.83333333333333,0.584705986078456)
--(axis cs:2.83333333333333,0.617091897483209);

\path [draw=black, semithick]
(axis cs:3.83333333333333,0.57161927969216)
--(axis cs:3.83333333333333,0.612014243357871);

\path [draw=black, semithick]
(axis cs:4.83333333333333,0.554351592419279)
--(axis cs:4.83333333333333,0.59537588266448);

\path [draw=black, semithick]
(axis cs:5.83333333333333,0.548867476131164)
--(axis cs:5.83333333333333,0.594925229303551);

\path [draw=black, semithick]
(axis cs:6.83333333333333,0.55462030928199)
--(axis cs:6.83333333333333,0.588578070959198);

\path [draw=black, semithick]
(axis cs:7.83333333333333,0.540423203330624)
--(axis cs:7.83333333333333,0.581834806328653);

\path [draw=black, semithick]
(axis cs:8.83333333333333,0.528844761833993)
--(axis cs:8.83333333333333,0.576909795038002);

\path [draw=black, semithick]
(axis cs:9.83333333333333,0.45757424442212)
--(axis cs:9.83333333333333,0.489915572821938);

\path [draw=black, semithick]
(axis cs:10.8333333333333,0.437557898211575)
--(axis cs:10.8333333333333,0.473503121161393);

\path [draw=black, semithick]
(axis cs:11.8333333333333,0.40038082882317)
--(axis cs:11.8333333333333,0.456705442677653);

\addplot [semithick, black, mark=-, mark size=2, mark options={solid}, only marks]
table {%
-0.166666666666667 0.621509672914657
0.833333333333333 0.599949249621053
1.83333333333333 0.590733918699181
2.83333333333333 0.584705986078456
3.83333333333333 0.57161927969216
4.83333333333333 0.554351592419279
5.83333333333333 0.548867476131164
6.83333333333333 0.55462030928199
7.83333333333333 0.540423203330624
8.83333333333333 0.528844761833993
9.83333333333333 0.45757424442212
10.8333333333333 0.437557898211575
11.8333333333333 0.40038082882317
};
\addplot [semithick, black, mark=-, mark size=2, mark options={solid}, only marks]
table {%
-0.166666666666667 0.669767556742262
0.833333333333333 0.639387935663913
1.83333333333333 0.626576986183954
2.83333333333333 0.617091897483209
3.83333333333333 0.612014243357871
4.83333333333333 0.59537588266448
5.83333333333333 0.594925229303551
6.83333333333333 0.588578070959198
7.83333333333333 0.581834806328653
8.83333333333333 0.576909795038002
9.83333333333333 0.489915572821938
10.8333333333333 0.473503121161393
11.8333333333333 0.456705442677653
};
\path [draw=black, semithick]
(axis cs:0.166666666666667,0.592996580197159)
--(axis cs:0.166666666666667,0.631870203427539);

\path [draw=black, semithick]
(axis cs:1.16666666666667,0.596619494527057)
--(axis cs:1.16666666666667,0.627599423371383);

\path [draw=black, semithick]
(axis cs:2.16666666666667,0.578685480333282)
--(axis cs:2.16666666666667,0.606542240696501);

\path [draw=black, semithick]
(axis cs:3.16666666666667,0.571040390332391)
--(axis cs:3.16666666666667,0.597200745549114);

\path [draw=black, semithick]
(axis cs:4.16666666666667,0.572236597534722)
--(axis cs:4.16666666666667,0.609541336782951);

\path [draw=black, semithick]
(axis cs:5.16666666666667,0.5729635484508)
--(axis cs:5.16666666666667,0.603287245698792);

\path [draw=black, semithick]
(axis cs:6.16666666666667,0.528723407044691)
--(axis cs:6.16666666666667,0.573119818826414);

\path [draw=black, semithick]
(axis cs:7.16666666666667,0.554954031241373)
--(axis cs:7.16666666666667,0.585241986424292);

\path [draw=black, semithick]
(axis cs:8.16666666666667,0.553653388603314)
--(axis cs:8.16666666666667,0.589387891585132);

\path [draw=black, semithick]
(axis cs:9.16666666666667,0.525523940331061)
--(axis cs:9.16666666666667,0.569120627849463);

\path [draw=black, semithick]
(axis cs:10.1666666666667,0.45346253131398)
--(axis cs:10.1666666666667,0.488331144467366);

\path [draw=black, semithick]
(axis cs:11.1666666666667,0.447569305488935)
--(axis cs:11.1666666666667,0.482671941489486);

\path [draw=black, semithick]
(axis cs:12.1666666666667,0.362243762388635)
--(axis cs:12.1666666666667,0.415451853416828);

\addplot [semithick, black, mark=-, mark size=2, mark options={solid}, only marks]
table {%
0.166666666666667 0.592996580197159
1.16666666666667 0.596619494527057
2.16666666666667 0.578685480333282
3.16666666666667 0.571040390332391
4.16666666666667 0.572236597534722
5.16666666666667 0.5729635484508
6.16666666666667 0.528723407044691
7.16666666666667 0.554954031241373
8.16666666666667 0.553653388603314
9.16666666666667 0.525523940331061
10.1666666666667 0.45346253131398
11.1666666666667 0.447569305488935
12.1666666666667 0.362243762388635
};
\addplot [semithick, black, mark=-, mark size=2, mark options={solid}, only marks]
table {%
0.166666666666667 0.631870203427539
1.16666666666667 0.627599423371383
2.16666666666667 0.606542240696501
3.16666666666667 0.597200745549114
4.16666666666667 0.609541336782951
5.16666666666667 0.603287245698792
6.16666666666667 0.573119818826414
7.16666666666667 0.585241986424292
8.16666666666667 0.589387891585132
9.16666666666667 0.569120627849463
10.1666666666667 0.488331144467366
11.1666666666667 0.482671941489486
12.1666666666667 0.415451853416828
};
\end{axis}

\end{tikzpicture}\\
			\vspace{-1mm}
			\setlength\figureheight{4.7cm}
			\setlength\figurewidth{18cm}
			\input{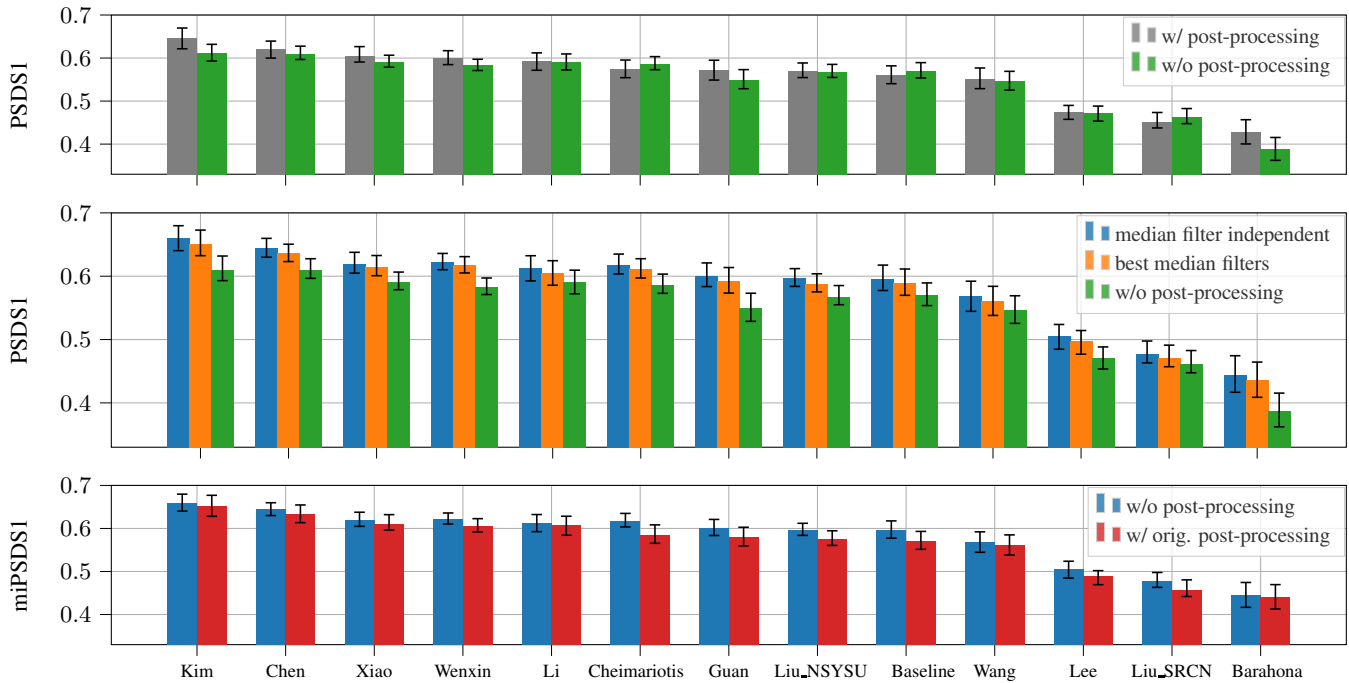}\\
			\vspace{-1mm}
			\setlength\figureheight{3.7cm}
			\setlength\figurewidth{18cm}
\begin{tikzpicture}

\definecolor{crimson2143940}{RGB}{214,39,40}
\definecolor{darkgray176}{RGB}{176,176,176}
\definecolor{lightgray204}{RGB}{204,204,204}
\definecolor{steelblue31119180}{RGB}{31,119,180}

\begin{axis}[
height=\figureheight,
width=\figurewidth,
legend cell align={left},
legend style={fill opacity=0.8, draw opacity=1, text opacity=1, draw=lightgray204,at={(.999,.99)}},
tick align=outside,
tick pos=left,
x grid style={darkgray176},
xmajorgrids,
xmin=-0.966666666666667, xmax=12.9666666666667,
xtick style={color=black},
xtick={0,1,2,3,4,5,6,7,8,9,10,11,12},
xticklabels={{\scriptsize Kim},{\scriptsize Chen},{\scriptsize Xiao},{\scriptsize Wenxin},{\scriptsize Li},{\scriptsize \hspace{-1mm}Cheimariotis},{\scriptsize \hspace{0mm}Guan},{\scriptsize \hspace{1mm}Liu\_NSYSU},{\scriptsize\hspace{4mm} Baseline},{\scriptsize \hspace{0mm}Wang},{\scriptsize Lee},{\scriptsize Liu\_SRCN},{ \hspace{1mm}\scriptsize Barahona}},
y grid style={darkgray176},
ylabel={miPSDS1},
ymajorgrids,
ymin=0.33, ymax=0.7,
ytick style={color=black}
]
\draw[draw=none,fill=steelblue31119180] (axis cs:-0.333333333333333,0) rectangle (axis cs:0,0.659260979980118);
\addlegendimage{ybar,ybar legend,draw=none,fill=steelblue31119180}
\addlegendentry{\footnotesize w/o post-processing}

\draw[draw=none,fill=steelblue31119180] (axis cs:0.666666666666667,0) rectangle (axis cs:1,0.6437715659882);
\draw[draw=none,fill=steelblue31119180] (axis cs:1.66666666666667,0) rectangle (axis cs:2,0.619295905167802);
\draw[draw=none,fill=steelblue31119180] (axis cs:2.66666666666667,0) rectangle (axis cs:3,0.621922500683363);
\draw[draw=none,fill=steelblue31119180] (axis cs:3.66666666666667,0) rectangle (axis cs:4,0.612188883096801);
\draw[draw=none,fill=steelblue31119180] (axis cs:4.66666666666667,0) rectangle (axis cs:5,0.617561102397129);
\draw[draw=none,fill=steelblue31119180] (axis cs:5.66666666666667,0) rectangle (axis cs:6,0.60075090958735);
\draw[draw=none,fill=steelblue31119180] (axis cs:6.66666666666667,0) rectangle (axis cs:7,0.596198290769851);
\draw[draw=none,fill=steelblue31119180] (axis cs:7.66666666666667,0) rectangle (axis cs:8,0.595015577963831);
\draw[draw=none,fill=steelblue31119180] (axis cs:8.66666666666667,0) rectangle (axis cs:9,0.567669077200743);
\draw[draw=none,fill=steelblue31119180] (axis cs:9.66666666666667,0) rectangle (axis cs:10,0.505102345212296);
\draw[draw=none,fill=steelblue31119180] (axis cs:10.6666666666667,0) rectangle (axis cs:11,0.476444635958079);
\draw[draw=none,fill=steelblue31119180] (axis cs:11.6666666666667,0) rectangle (axis cs:12,0.444289446131666);
\draw[draw=none,fill=crimson2143940] (axis cs:-2.77555756156289e-17,0) rectangle (axis cs:0.333333333333333,0.651850445089085);
\addlegendimage{ybar,ybar legend,draw=none,fill=crimson2143940}
\addlegendentry{\footnotesize w/ orig. post-processing}

\draw[draw=none,fill=crimson2143940] (axis cs:1,0) rectangle (axis cs:1.33333333333333,0.633909522184436);
\draw[draw=none,fill=crimson2143940] (axis cs:2,0) rectangle (axis cs:2.33333333333333,0.610452404092179);
\draw[draw=none,fill=crimson2143940] (axis cs:3,0) rectangle (axis cs:3.33333333333333,0.605859497454285);
\draw[draw=none,fill=crimson2143940] (axis cs:4,0) rectangle (axis cs:4.33333333333333,0.606622997346799);
\draw[draw=none,fill=crimson2143940] (axis cs:5,0) rectangle (axis cs:5.33333333333333,0.584655210763009);
\draw[draw=none,fill=crimson2143940] (axis cs:6,0) rectangle (axis cs:6.33333333333333,0.580398662994272);
\draw[draw=none,fill=crimson2143940] (axis cs:7,0) rectangle (axis cs:7.33333333333333,0.576071325663127);
\draw[draw=none,fill=crimson2143940] (axis cs:8,0) rectangle (axis cs:8.33333333333333,0.570714961849322);
\draw[draw=none,fill=crimson2143940] (axis cs:9,0) rectangle (axis cs:9.33333333333333,0.560059522160319);
\draw[draw=none,fill=crimson2143940] (axis cs:10,0) rectangle (axis cs:10.3333333333333,0.488134273453521);
\draw[draw=none,fill=crimson2143940] (axis cs:11,0) rectangle (axis cs:11.3333333333333,0.457053429793059);
\draw[draw=none,fill=crimson2143940] (axis cs:12,0) rectangle (axis cs:12.3333333333333,0.439407842268456);
\path [draw=black, semithick]
(axis cs:-0.166666666666667,0.640311523726675)
--(axis cs:-0.166666666666667,0.679714902026828);

\path [draw=black, semithick]
(axis cs:0.833333333333333,0.629972219373758)
--(axis cs:0.833333333333333,0.659756516256228);

\path [draw=black, semithick]
(axis cs:1.83333333333333,0.604841869394773)
--(axis cs:1.83333333333333,0.637724275559517);

\path [draw=black, semithick]
(axis cs:2.83333333333333,0.610202072506399)
--(axis cs:2.83333333333333,0.63596195297884);

\path [draw=black, semithick]
(axis cs:3.83333333333333,0.592587690044005)
--(axis cs:3.83333333333333,0.632328295197982);

\path [draw=black, semithick]
(axis cs:4.83333333333333,0.603608441304357)
--(axis cs:4.83333333333333,0.634943584349368);

\path [draw=black, semithick]
(axis cs:5.83333333333333,0.58361820973589)
--(axis cs:5.83333333333333,0.620986654940399);

\path [draw=black, semithick]
(axis cs:6.83333333333333,0.583848014983202)
--(axis cs:6.83333333333333,0.611986680839788);

\path [draw=black, semithick]
(axis cs:7.83333333333333,0.577453877868864)
--(axis cs:7.83333333333333,0.61752683863735);

\path [draw=black, semithick]
(axis cs:8.83333333333333,0.544593436374341)
--(axis cs:8.83333333333333,0.592110623782618);

\path [draw=black, semithick]
(axis cs:9.83333333333333,0.484667910027769)
--(axis cs:9.83333333333333,0.523807191620263);

\path [draw=black, semithick]
(axis cs:10.8333333333333,0.463150938846514)
--(axis cs:10.8333333333333,0.497684318111853);

\path [draw=black, semithick]
(axis cs:11.8333333333333,0.416786577392531)
--(axis cs:11.8333333333333,0.474468957759443);

\addplot [semithick, black, mark=-, mark size=2, mark options={solid}, only marks]
table {%
-0.166666666666667 0.640311523726675
0.833333333333333 0.629972219373758
1.83333333333333 0.604841869394773
2.83333333333333 0.610202072506399
3.83333333333333 0.592587690044005
4.83333333333333 0.603608441304357
5.83333333333333 0.58361820973589
6.83333333333333 0.583848014983202
7.83333333333333 0.577453877868864
8.83333333333333 0.544593436374341
9.83333333333333 0.484667910027769
10.8333333333333 0.463150938846514
11.8333333333333 0.416786577392531
};
\addplot [semithick, black, mark=-, mark size=2, mark options={solid}, only marks]
table {%
-0.166666666666667 0.679714902026828
0.833333333333333 0.659756516256228
1.83333333333333 0.637724275559517
2.83333333333333 0.63596195297884
3.83333333333333 0.632328295197982
4.83333333333333 0.634943584349368
5.83333333333333 0.620986654940399
6.83333333333333 0.611986680839788
7.83333333333333 0.61752683863735
8.83333333333333 0.592110623782618
9.83333333333333 0.523807191620263
10.8333333333333 0.497684318111853
11.8333333333333 0.474468957759443
};
\path [draw=black, semithick]
(axis cs:0.166666666666667,0.628167639001822)
--(axis cs:0.166666666666667,0.677001148611493);

\path [draw=black, semithick]
(axis cs:1.16666666666667,0.61347730010678)
--(axis cs:1.16666666666667,0.654603747931438);

\path [draw=black, semithick]
(axis cs:2.16666666666667,0.596332794973994)
--(axis cs:2.16666666666667,0.632080440881281);

\path [draw=black, semithick]
(axis cs:3.16666666666667,0.591919323881618)
--(axis cs:3.16666666666667,0.622770982346615);

\path [draw=black, semithick]
(axis cs:4.16666666666667,0.584228803603512)
--(axis cs:4.16666666666667,0.628354805993682);

\path [draw=black, semithick]
(axis cs:5.16666666666667,0.565507832084354)
--(axis cs:5.16666666666667,0.608331412090225);

\path [draw=black, semithick]
(axis cs:6.16666666666667,0.559432871370824)
--(axis cs:6.16666666666667,0.602774575119652);

\path [draw=black, semithick]
(axis cs:7.16666666666667,0.560543717818845)
--(axis cs:7.16666666666667,0.594582412810297);

\path [draw=black, semithick]
(axis cs:8.16666666666667,0.551487324175624)
--(axis cs:8.16666666666667,0.593116326256582);

\path [draw=black, semithick]
(axis cs:9.16666666666667,0.538369137119836)
--(axis cs:9.16666666666667,0.585041086700774);

\path [draw=black, semithick]
(axis cs:10.1666666666667,0.469348018001583)
--(axis cs:10.1666666666667,0.501960080716489);

\path [draw=black, semithick]
(axis cs:11.1666666666667,0.441829722793682)
--(axis cs:11.1666666666667,0.480518945282597);

\path [draw=black, semithick]
(axis cs:12.1666666666667,0.412738428747385)
--(axis cs:12.1666666666667,0.469503910314934);

\addplot [semithick, black, mark=-, mark size=2, mark options={solid}, only marks]
table {%
0.166666666666667 0.628167639001822
1.16666666666667 0.61347730010678
2.16666666666667 0.596332794973994
3.16666666666667 0.591919323881618
4.16666666666667 0.584228803603512
5.16666666666667 0.565507832084354
6.16666666666667 0.559432871370824
7.16666666666667 0.560543717818845
8.16666666666667 0.551487324175624
9.16666666666667 0.538369137119836
10.1666666666667 0.469348018001583
11.1666666666667 0.441829722793682
12.1666666666667 0.412738428747385
};
\addplot [semithick, black, mark=-, mark size=2, mark options={solid}, only marks]
table {%
0.166666666666667 0.677001148611493
1.16666666666667 0.654603747931438
2.16666666666667 0.632080440881281
3.16666666666667 0.622770982346615
4.16666666666667 0.628354805993682
5.16666666666667 0.608331412090225
6.16666666666667 0.602774575119652
7.16666666666667 0.594582412810297
8.16666666666667 0.593116326256582
9.16666666666667 0.585041086700774
10.1666666666667 0.501960080716489
11.1666666666667 0.480518945282597
12.1666666666667 0.469503910314934
};
\end{axis}

\end{tikzpicture}
			\vspace{-6mm}
			\caption{System Evaluation. Upper Plot: original post-processing vs. no post-processing. Middle Plot: \gls{miPSDS} vs. \gls{PSDS} with optimal median filter lengths per class vs. no post-processing. Lower Plot: miPSDS computed with unprocessed vs. post-processed data. }
			\label{fig:system_eval}
			\vspace{-4mm}
		\end{figure*}

		We first run investigations on the baseline system Baseline\_BEATS~\cite{baseline2023beats} (Baseline).
		In the upper subplot of Fig.~\ref{fig:baseline_rocs} we compare the following PSD-ROCs:
        \vspace{-1mm}
		\begin{enumerate}
			\itemsep0mm 
			\item median filter independent: as defined in Eq.~\ref{eq:piPSDROC}
			\item best median filters: choosing best performing median filter per class as follows\\
			$\tilde{r}_c(e) = r_{c,b}(e)\text{ with }b = \mathrm{argmax}_l{\mathrm{auc}(r_{c,l}(e))}$,
			\item without any post-processing.
		\end{enumerate}
        \vspace{-1mm}
		It can be seen, that by applying (best) median filtering the \gls{PSDROC} can be significantly improved over the unprocessed case.
		It can be further observed, that there are operating points, especially for low \glspl{eFPR}, where the \gls{miPSDROC} is higher than best median filter PSD-ROC.
		This indicates that best median filters are, although giving best overall performance, not the best choice for each individual operating point and better performance can be achieved by choosing operating point dependent filter lengths as the \gls{miPSDROC} does.
        In the lower subplot of Fig.~\ref{fig:baseline_rocs} we plot, for some event classes, the optimal filter lengths over operating points.
		We can see that for lower \glspl{eFPR} optimal median filters tend to be longer than for higher \glspl{eFPR}, which can be explained by the fact that longer median filters better suppress short duration \glspl{FP}.
		Further, event classes with longer per-event durations, such as "Frying" and "Running Water", tend to have overall longer median filters than short duration event classes, which makes intuitively sense.

		Next, we evaluate challenge submissions\footnote{Submissions will be made publicly available on zenodo soon with link being added in the camera-ready version in case of acceptance} with, without and independent of post-processing.
		As submitting unprocessed scores was optional, we evaluate only systems from the 12 teams that did provide them.
		We limit evaluation to the one single-model system per team that gave best PSDS1 performance in the challenge (with original post-processing).
		These systems are Barahona-AUDIAS-2~\cite{Barahona2023}, Cheimariotis-DUTH-1~\cite{Cheimariotis2023}, Chen-CHT-2~\cite{Chen2023}, Guan-HIT-3~\cite{Guan2023}, Kim-GIST-HanwhaVision-2~\cite{Kim2023}, Lee-CAUET-1~\cite{Lee2023}, Li-USTC-6~\cite{Li2023}, Liu-NSYSU-7~\cite{LiuNSYSU2023}, Liu-SRCN-4~\cite{LiuSRCN2023a},  Wang-XiaoRice-1~\cite{Wang2023}, Wenxin-TJU-6~\cite{Wenxin2023}, Xiao-FMSG-4~\cite{Xiao2023}.

		To be able to evaluate the variance of system performance over different runs of system training, participants submitted prediction scores for three runs of training for each system.
		To further track variance of results due to variations in the evaluation data, we perform bootstrapped evaluation, where evaluation is performed on 20 different \SI{80}{\%} fractions of the eval data.
		In total we evaluate $3\cdot20=60$ different setups and report the mean and  $\SI{5}{\%}-\SI{95}{\%}$ confidence interval of the system's performances.
		This evaluation procedure is the same as we used for official challenge evaluation.

		We first want to investigate the impact of the post-processing on the systems' performances in the upper subplot of Fig.~\ref{fig:system_eval}. by comparing the performance with and without the post-processing as used by the participants.
		It appears that for some systems, e.g.,  Kim-GIST-HanwhaVision-2, the performance significantly degrades when removing the post-processing, whereas for other systems the performance does not degrade or even improves.
		When evaluating the unprocessed scores, the ranking also changes at multiple positions to Kim, Chen, Li, Xiao, Cheimariotis, Wenxin, Baseline, Liu\_NSYSU, Guan, Wang, Lee, Liu\_SRCN, Barahona. 
		This suggests that there is some bias introduced by the post-processing, particularly, whether a sophisticated post-processing is employed or not. 
		To some extent, however, it may also be a system property that it can benefit from post-processing more than other systems.

		We next evaluate our proposed \gls{miPSDS} and compare it to "no processing" and "best median filters"  in the middle subplot of Fig.~\ref{fig:system_eval}.
		It can be seen that for all systems performance can be improved by best median filters and further improved by operating point specific median filters as considered by \gls{miPSDS}.
		Some systems, e.g., Kim and Barahona, benefit more from best median filters / median filter independent evaluation than others, which can be explained by our previous assumption that the effectiveness of post-processing is to some extent also a system property.
		Here, \gls{miPSDS} evaluation gives again a different ranking which is Kim, Chen, Wenxin, Xiao, Cheimariotis, Li, Guan, Liu\_NSYSU, Baseline, Wang, Lee, Liu\_SRCN, Barahon.
		
		Note, that it is still possible to run additional post-processing before \gls{piPSDS} evaluation to improve performance.
		However, it can be assumed that the possible gain is rather small and it is more likely that an additional post-processing degrades \gls{piPSDS}.
		To investigate this, we compare \gls{miPSDS} evaluated on unprocessed scores vs. scores with participants' original post-processing in the lower subplot of Fig.~\ref{fig:system_eval}.
		It can be seen that in all cases the additional post-processing degrades \gls{miPSDS} performance.

        \vspace{-1mm}
		\section{Conclusions}
        \vspace{-2mm}
		\label{sec:conclusions}
		Due to the high variation of \gls{SED} system application requirements, \gls{SED} evaluation has to capture the overall system behavior over various operating points.
		Therefore, the community recently moved to decision threshold independent evaluation using \glspl{PSDS} to capture performance over different decision thresholds used for binarization of system output scores.
		In this paper we proposed \gls{piPSDS} which further evaluates performance over different post-processings and effectively choosing the post-processing that is best suited for a certain operating mode.
		It has been shown that \gls{piPSDS} indeed overcomes the bias introduced due to different post-processings but still accounts for system-specific effectiveness of post-processing.
		It further allows for system comparison without the need of employing a sophisticated post-processing, e.g., during system development.

        \balance
		\bibliographystyle{IEEEtran}
		\bibliography{refs}
		
		%
			%
			%
			%
			%
			%
			%
			%
			%

	\end{sloppy}
\end{document}